\documentclass{ieeeaccess}
\usepackage[T1]{fontenc}
\usepackage{lmodern}
\usepackage[english]{babel}
\usepackage{booktabs}
\usepackage{enumitem}
\usepackage{hyperref}
\usepackage{xspace}
\usepackage{array}
\usepackage{ragged2e}
\usepackage{listings}
\usepackage[leftmargin=0.5cm, rightmargin=0.5cm, vskip=6pt]{quoting}
\usepackage[autostyle=true]{csquotes}
\usepackage{balance}
\usepackage{algorithm}
\usepackage{algpseudocode}

\usepackage[numbers]{natbib}
\usepackage{amsmath}
\usepackage{amssymb}
\usepackage{amsthm}
\usepackage{amsfonts}

\usepackage{graphicx}
\usepackage{caption}
\usepackage{needspace}
\usepackage{tabularx}

\newcommand{\inlinefigpage}[4]{%
\begin{figure}[!t]
\centering
\includegraphics[page=#1,width=0.96\columnwidth]{figures.pdf}
\caption{#2}
\label{#3}
\end{figure}
}
\newcommand{\cpntwo}{%
  \texorpdfstring{\textsc{CP}\textsuperscript{2}\textsc{N}\textsuperscript{2}}%
  {CP2N2}\xspace}

\SetBlockEnvironment{quoting}
\def\BibTeX{{\rm B\kern-.05em{\sc i\kern-.025em b}\kern-.08em
    T\kern-.1667em\lower.7ex\hbox{E}\kern-.125emX}}
\newcolumntype{C}{>{\centering\arraybackslash}X}
\newcolumntype{R}{>{\raggedleft\arraybackslash}X}

\begin{document}
\history{Date of publication xxxx 00, 0000, date of current version xxxx 00, 0000.}
\doi{10.1109/ACCESS.2017.DOI}

\newcommand{\myorcid}[1]{%
  \href{https://orcid.org/#1}{\includegraphics[width=8pt]{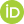}}}

\title{Trust-Aware Output Management for Physical Neural Network in Cloud-Continuum Systems}
\author{
    \uppercase{Maliheh Hariri}\authorrefmark{1}\,\myorcid{0000-0002-1700-8637},
    \uppercase{Stefan Fischer}\authorrefmark{2}\,\myorcid{0000-0003-1292-8925},
    \IEEEmembership{Member, IEEE}
}

\address[1, 2]{University of L\"ubeck, Institute of Telematics, 23562 L\"ubeck, Germany}

\markboth
{Hariri \headeretal: Trust-Aware Management of Physical
Neural Network Outputs}
{Hariri \headeretal: Trust-Aware Management of Physical
Neural Network Outputs}

\corresp{Corresponding author: Stefan Fischer (e-mail: stefan.fischer@uni-luebeck.de).}

\begin{abstract}

Physical Neural Networks (PNNs) introduce new opportunities for cloud continuum computing, but their outputs may be affected by noise, drift, delay, and incomplete reliability information. Existing substrate-management approaches mainly focus on discovery, invocation, and monitoring, while the reliability of the returned output is often left unaddressed. This paper proposes a trust-aware output management framework for heterogeneous PNNs. Each output is represented with quality and context information, and a lightweight edge-level trust score decides whether it should be accepted, rejected, or forwarded to the fog. At the fog layer, compatible outputs are checked for disagreement and combined using reliability- and uncertainty-aware fusion. Historical trust is also tracked to detect sustained degradation and support recalibration requests.
The framework is evaluated using controlled and randomized PNN output models. Across 20 random seeds, the proposed full-trust policy reduces unsafe acceptance from approximately 61.5\% for raw output handling to about 5.0\%, while accepted-output MAE decreases from 0.504 to 0.242. Risk coverage analysis shows that this improvement is not explained only by lower local acceptance. The proposed fog fusion method also achieves the lowest aggregate mean error among the evaluated methods, with a small but consistent advantage over strong uncertainty-aware baselines. The prototype adds about $2~\mu\text{s}$ of edge processing per evidence record. These results show that post-invocation reliability management can improve the safe use of PNN outputs across edge--fog--cloud systems.
\end{abstract}


\titlepgskip = -15pt

\maketitle

\section{Introduction}
\label{sec:introduction}
Cloud continuum systems distribute computation across edge, fog, and cloud resources instead of relying only on centralized infrastructure. This supports low-latency processing and closer interaction with the physical environment. As these systems evolve, they are also beginning to include non-conventional computing resources in addition to digital processors and accelerators.

Physical Neural Networks (PNNs) are one example. They use the dynamics of physical substrates, including chemical, biological, memristive, photonic, and other material systems, as part of the computation itself~\cite{fischer2026beyondsilicon}. Previous work has shown that nonlinear responses, memory effects, and device dynamics can provide useful computational behavior~\cite{tanaka2019physicalreservoir,markovic2020physics}. These properties make PNNs relevant for edge and extreme-edge environments, but they also make their outputs sensitive to physical conditions.

A PNN output may be affected by noise, drift, delay, calibration state, aging, or environmental changes. Its result may also take different forms, such as spike activity, optical intensity, conductance, or chemical concentration. Therefore, successful execution alone does not guarantee that the returned output is reliable enough to use in a distributed decision process.

Several systems support the programming and access of physical computing substrates. NIR and EdgeMap improve portability and deployment for neuromorphic systems~\cite{pedersen2024nir,xue2023edgemap}, while BioCRNpyler supports compilation for chemical reaction networks~\cite{poole2022biocrnpyler}. Wetware platforms such as DishBrain, NeuroPlatform, and the Cortical Labs API provide software-facing access to biological neural systems~\cite{kagan2022dishbrain,jordan2024neuroplatform,hogan2026clapi}. More recently, \cpntwo provides a substrate-aware control plane for discovering, invoking, and monitoring heterogeneous PNN resources across the cloud continuum~\cite{fischer2026phys}. Related ideas also appear in the Model Context Protocol, Web of Things, and digital-twin systems~\cite{mcp2025spec,w3c2023wotarchitecture,kamburjan2024declarative}.

These approaches mainly focus on how a substrate is described, selected, invoked, or monitored. They do not explicitly decide whether the output returned after invocation should be trusted. This becomes important for PNNs because a correctly invoked substrate may still produce an untrustable result. The problem considered in this paper is therefore the reliability of the output after successful invocation.

To address this problem, we introduce a trust-aware output management framework for PNNs in the cloud continuum. The main contributions are:

\begin{itemize}
    \item A common evidence representation and edge-level trust model that considers confidence, uncertainty, drift, and freshness when deciding whether an output should be accepted, rejected, or forwarded.

    \item A fog-level mechanism that checks compatibility and source disagreement before applying reliability- and uncertainty-aware fusion to eligible PNN outputs.

    \item An adaptive trust mechanism that tracks reliability over time and can trigger recalibration requests when persistent degradation is detected.
\end{itemize}

The remainder of the paper is organized as follows. Section~\ref{sec:relatedworks} reviews related work. Section~\ref{sec:SystemModel} presents the system model and assumptions. Section~\ref{sec:method} describes the proposed framework. Section~\ref{sec:eval} presents the evaluation, and Section~\ref{sec:conclusion} concludes the paper.

\section{Related Work}
\label{sec:relatedworks}

The reliability of Physical Neural Network (PNN) outputs is related to several areas, including physical neural computing, edge intelligence, uncertainty-aware inference, sensor fusion, trustworthy edge computing, and adaptive system management. These areas provide useful methods for handling heterogeneous computation and uncertain data, but none directly addresses the post-invocation question considered in this work: whether a returned PNN output is reliable enough to be accepted, rejected, or further processed.

Fischer et al.~\cite{fischer2026beyondsilicon} review physical neural computing across memristive, photonic, chemical, mechanical, and other substrates, showing both the computational potential and the strong dependence of these systems on their physical dynamics. Tanaka et al.~\cite{tanaka2019physicalreservoir} and Markovi\'c et al.~\cite{markovic2020physics} similarly show how physical dynamics can provide efficient computation, but also make performance dependent on device and environmental conditions; these works study the computing mechanisms rather than how their outputs should be trusted after execution.

Several works instead focus on making heterogeneous substrates easier to program or access. Pedersen et al.~\cite{pedersen2024nir} introduce a common intermediate representation for neuromorphic systems, while Xue et al.~\cite{xue2023edgemap} optimize the mapping of spiking neural networks to edge hardware. These approaches improve portability and deployment, but assume that a successfully executed model produces an output that can be used directly.

Similar abstractions exist for other physical substrates. BioCRNpyler~\cite{poole2022biocrnpyler} and ChemComp~\cite{agostini2025chemcomp} provide compilation methods for chemical reaction networks, reducing the gap between high-level models and physical implementations. Their focus is on translating and executing computations, not on judging the reliability of the physical result returned after execution.

Wetware platforms make this issue even more visible. Kagan et al.~\cite{kagan2022dishbrain} demonstrated closed-loop interaction with cultured neural cells, while Jordan et al.~\cite{jordan2024neuroplatform} and Hogan et al.~\cite{hogan2026clapi} provide remote and software-facing access to biological neural systems. These platforms show that physical neural computation can be exposed through software interfaces, but output quality remains closely tied to biological state and experimental conditions and is not handled through a common post-invocation trust mechanism.

Shi et al.~\cite{shi2020communication} study communication-efficient Edge AI and show how distributing learning and inference closer to data sources can reduce latency and communication cost. Xu et al.~\cite{xu2021edge} provide a broader view of edge intelligence in which data, models, computation, and decisions are distributed across the network; both approaches motivate local decision-making but mainly treat model outputs as conventional digital results.

Murshed et al.~\cite{murshed2021ml_edge} review machine-learning deployment at the network edge, including model compression, hardware support, and resource constraints. Baccour et al.~\cite{baccour2022pervasive} focus on resource-efficient distributed AI for IoT systems and consider computation, communication, and energy jointly. These works provide the systems basis for edge--cloud execution, but do not evaluate the reliability of a physical output after the underlying computation has completed.

Qendro et al.~\cite{qendro2021benefit} introduce a lightweight uncertainty-aware sensing approach suitable for resource-constrained edge devices. Their method makes prediction uncertainty available with limited overhead, but mainly captures uncertainty of a digital learning model rather than physical effects.

This distinction matters for PNNs because uncertainty can originate from both computation and the physical substrate itself. A returned value may therefore appear confident at the model level while still being affected by changing physical conditions, which motivates combining several reliability indicators rather than relying on predictive confidence alone.

Gruber et al.~\cite{gruber2021uncertainty} show that uncertainty-aware sensor fusion can reduce the influence of unreliable measurements compared with treating all sensor values equally. Their later work~\cite{gruber2022application} applies this idea to physical sensor networks and connects uncertainty information with digital representations of sensing entities. These methods provide a useful basis for weighted fusion, but they mainly consider measurements from conventional sensors rather than computational outputs produced by heterogeneous physical neural substrates.

Vedurmudi et al.~\cite{vedurmudi2025automation} review automated uncertainty handling in sensor-network metrology, including middleware, fusion, machine learning, and agent-based techniques. This work treats uncertainty as part of the data-processing pipeline, but generally assumes sensing systems with defined measurement and calibration models; PNN outputs may additionally depend on the internal dynamics of the computing substrate.

Wang et al.~\cite{wang2024trustworthy} review trustworthy edge intelligence from the perspectives of reliability, security, transparency, and sustainability. Their work provides a broad view of trust in distributed AI systems, but does not define an output-level decision process for noisy, drifting, or stale physical neural computations.

Hallyburton et al.~\cite{hallyburton2025mate} use adaptive trust estimation in sensor fusion, allowing the influence of a source to change as new observations become available. This supports the idea that reliability should evolve over time, although their setting mainly addresses security and faulty sensing rather than degradation caused by the physical computing substrate itself.

Tan and Matta~\cite{tan2024dtsync} study the synchronization problem between physical systems and their digital twins and formalize when twin state should be updated. Kamburjan et al.~\cite{kamburjan2024declarative} address lifecycle management of digital twins through explicit models of system state and evolution. These approaches are useful for long-term monitoring and recovery, but they do not provide a mechanism for deciding whether an individual PNN output should be trusted immediately after invocation.

We summarize all research directions in Table~\ref{tab:related_work_comparison} which are mainly address conventional digital AI models, sensor measurements, security-aware fusion, or substrate-level control. None of the proposed papers directly address the post-invocation reliability problem of heterogeneous PNN outputs.

\begin{table*}[t]
\centering
\caption{Comparison of the proposed framework with related research directions.}
\label{tab:related_work_comparison}
\scriptsize
\setlength{\tabcolsep}{5pt}
\begin{tabular}{p{5.0cm}ccccc}
\hline
\textbf{Research direction} &
\textbf{Edge/Fog} &
\textbf{Uncertainty} &
\textbf{Fusion/Trust} &
\textbf{PNN-aware} &
\textbf{Post-invocation output trust} \\
\hline

Physical neural computing and reservoir systems
\cite{fischer2026beyondsilicon,tanaka2019physicalreservoir,markovic2020physics}
& -- & \textit{Partial} & -- & \checkmark & -- \\

Substrate programming and deployment
\cite{pedersen2024nir,xue2023edgemap,poole2022biocrnpyler,agostini2025chemcomp}
& \textit{Partial} & -- & -- & \checkmark & -- \\

Wetware access and biological computing
\cite{kagan2022dishbrain,jordan2024neuroplatform,hogan2026clapi}
& \textit{Partial} & -- & -- & \checkmark & -- \\

Edge intelligence
\cite{shi2020communication,xu2021edge,murshed2021ml_edge,baccour2022pervasive}
& \checkmark & -- & -- & -- & -- \\

Uncertainty-aware edge inference
\cite{qendro2021benefit}
& \checkmark & \checkmark & -- & -- & -- \\

Uncertainty-aware sensor fusion and metrology
\cite{gruber2021uncertainty,gruber2022application,vedurmudi2025automation}
& \checkmark & \checkmark & \checkmark & -- & -- \\

Trustworthy edge AI and adaptive trust
\cite{wang2024trustworthy,hallyburton2025mate}
& \checkmark & \checkmark & \checkmark & -- & -- \\

Digital-twin synchronization and lifecycle management
\cite{tan2024dtsync,kamburjan2024declarative}
& \textit{Partial} & \textit{Partial} & -- & -- & -- \\

\cpntwo{} / substrate control
\cite{fischer2026phys,mcp2025spec}
& \checkmark & \textit{Partial} & \textit{Partial} & \checkmark & -- \\

\textbf{Proposed framework}
& \checkmark & \checkmark & \checkmark & \checkmark & \checkmark \\
\hline

\end{tabular}
\end{table*}

\section{System Model}\label{sec:SystemModel}
\subsection{Assumptions}
We consider a cloud-continuum architecture with edge, fog, and cloud layers. PNN substrates are located close to the physical environment, mainly at the extreme edge or edge layer. Fog nodes provide local coordination among multiple substrates, while the cloud or management layer maintains longer-term information such as calibration history, historical trust, and digital-twin state.

PNN outputs are not assumed to be fully reliable digital values. Depending on the substrate and operating conditions, an output may be affected by noise, drift, delay, calibration error, or changes in the physical state. Each returned output is therefore represented as an evidence record containing its value together with available quality and context information, including confidence, uncertainty, drift, timestamp, modality, and provenance. These indicators are obtained from substrate adapters, runtime telemetry, calibration records, repeated observations, or digital-twin information rather than being manually supplied by the user.

The edge layer acts as the first reliability decision point after a PNN invocation. It performs lightweight trust estimation and decides whether an output should be accepted locally, rejected, or forwarded to the fog. When required quality information is missing, the output is treated conservatively and can be forwarded for further assessment.

The fog layer may receive outputs from several PNN sources referring to the same task or observation. Fusion is performed only when the evidence is compatible in task, modality, time, and decision context. Compatible outputs are checked for disagreement before numerical fusion, so strongly inconsistent evidence is not combined automatically.

The cloud or management layer keeps longer-term reliability information and may receive recalibration requests when sustained degradation is detected. The current prototype evaluates trust tracking and recalibration triggering, but does not directly modify physical calibration parameters or automatically synchronize a deployed digital twin.

Communication between continuum layers is assumed to have non-zero delay and limited bandwidth. Network optimization is outside the scope of this work; communication delay is reflected through evidence freshness, while compact evidence records are preferred over transferring raw substrate traces. The framework is intended for heterogeneous PNN technologies, including wetware, chemical, memristive, photonic, and other physical computing substrates, without assuming a common internal implementation.
\subsection{Problem Formulation}
Given a PNN output generated by a physical substrate, the objective is to decide whether the output is reliable enough for local use, should be rejected, or should be forwarded to the fog layer for additional fusion. Each output is represented as an evidence record containing the returned value and its quality indicators. The decision problem is therefore defined over an evidence record \(r_i\), where the system must select an action \(D_i \in \{\text{Accept}, \text{Reject}, \text{Forward}\}\).

The objective is to reduce unsafe acceptance of unreliable physical outputs while preserving useful local decision-making for reliable outputs. This creates a trade-off between reliability and local availability. Accepting too many outputs at the edge may propagate noisy, stale, or drifting physical results, whereas a more conservative policy reduces local coverage and increases the number of outputs forwarded for additional processing. The proposed framework addresses this trade-off through trust-aware edge scoring and fog-level compatibility checking, disagreement assessment, and fusion.

The trust score $T_i$ therefore serves as a compact reliability estimate derived from the quality indicators associated with $r_i$. Rather than being treated as an optimization objective itself, $T_i$ provides the basis for the threshold-based decision policy introduced in Section~\ref{sec:edge-decision}. The acceptance and rejection thresholds determine how conservatively the system operates and can be selected according to the reliability requirements and risk tolerance of the target application.

\section{Methodology}\label{sec:method}

The proposed methodology consists of five main stages: evidence construction and parameter estimation, edge-level trust computation, trust-based decision-making, fog-level compatibility checking and fusion, and adaptive feedback for long-term reliability management. The following subsections describe each stage in detail.

\subsection{Evidence Construction}

When a PNN substrate produces an output, the first step is to convert the raw physical signal into a structured evidence record. This step is required because PNNs may return different types of outputs depending on the substrate. 
%For example, a chemical PNN may return a concentration value, a wetware PNN may return spike activity, an optical system may return light intensity, and a memristive substrate may return a conductance or vector response. 
Although these outputs are physically different, the proposed framework represents them through a common evidence structure:

\begin{equation}
r_i = (v_i, c_i, \sigma_i, d_i, t_i, m_i, \rho_i),
\label{evidence}
\end{equation}

where $v_i$ denotes the output value returned by substrate $i$;
$c_i \in [0,1]$ is its confidence score;
$\sigma_i \in [0,1]$ represents the normalized noise or uncertainty level;
$d_i \in [0,1]$ represents the normalized drift from calibrated behavior;
$t_i$ is the output-generation timestamp;
$m_i$ identifies the output modality; and
$\rho_i$ contains provenance metadata, such as the substrate, adapter,
calibration, location, or digital-twin identifier.

The evidence record therefore captures both the physical output and the
information required to assess its reliability. Outputs with similar values
may consequently receive different trust assessments depending on their
confidence, noise, drift, freshness, and provenance.

\subsection{Parameter Estimation and Normalization}

The quality parameters in the evidence record  (Formula \ref{evidence} ) are not assumed to be manually provided by the user. They are estimated from telemetry, repeated measurements, adapter metadata, calibration records, timestamps, and digital-twin state. This makes the proposed method suitable for integration with substrate-aware control layers that already expose runtime and life cycle information.

The confidence value ($c_i$) in formula \ref{evidence} represents how reliable the output appears from the viewpoint of the substrate or adapter. It can be derived from signal strength, backend health, convergence quality, viability status, classification margin, or the stability of repeated responses. For example, a spike-based output with a clear and repeatable response pattern receives a higher confidence value than a weak or inconsistent response. The confidence score is normalized to the range ([0,1]), where 1 indicates high confidence, and 0 indicates very low confidence.

The uncertainty or noise value ($\sigma_i$) captures short-term instability in the output. If repeated measurements or samples are available, it can be estimated using the standard deviation of the returned values:

\begin{equation}
\sigma_i^{raw} =
\sqrt{\frac{1}{K}\sum_{k=1}^{K}(v_{ik}-\mu_i)^2},
\end{equation}

where ($v_{ik}$) is the (k)-th repeated output from substrate (i), (K) is the number of samples, and ($\mu_i$) is their mean. The raw uncertainty is then normalized as

\begin{equation}
\sigma_i = \min \left(\frac{\sigma_i^{raw}}{\sigma_i^{max}}, 1 \right),
\end{equation}

where ($\sigma_i^{max}$) is the maximum acceptable uncertainty for the corresponding task or substrate class. A lower value of ($\sigma_i$) means that the output is more stable.

The drift value ($d_i$) measures how far the current behavior of the substrate has moved away from its calibrated or expected behavior. This can be estimated by comparing the current response with a calibration baseline or with a digital-twin prediction:

\begin{equation}
d_i = \min \left(\frac{\Delta_i^{current}}{\Delta_i^{max}}, 1 \right),
\end{equation}

where ($\Delta_i^{current}$) is the deviation between the current response and the reference behavior, and ($\Delta_i^{max}$) is the maximum tolerable deviation. A small drift value means that the substrate is still close to its calibrated behavior, while a high drift value indicates that recalibration or reduced trust may be required.

Freshness is computed from the timestamp of the output. An output that was produced recently is usually more relevant than an old output, especially in time-sensitive edge applications. The freshness value is defined as

\begin{equation}
f_i = e^{-\Delta t_i/\tau},
\label{eq:freshness}
\end{equation}

where ($\Delta t_i = t_{now} - t_i$) is the age of the output and ($\tau$) is a task-dependent time constant. A larger ($\tau$) can be used for slow physical processes, while a smaller ($\tau$) is suitable for fast edge decisions.

If required reliability telemetry is missing, the framework applies a conservative decision override. Diagnostic estimates may still be constructed internally to preserve a complete evidence representation, but an output with required quality information missing is not accepted locally solely on the basis of these estimates. Instead, the evidence is marked as incomplete and forwarded for additional processing. This policy prevents an apparently high numerical trust score from masking the absence of information required to verify the reliability of the output.

\subsection{Edge-Level Trust Estimation}\label{sec:edge-decision}
After construction and normalization of evidence, the edge layer computes a trust score for each PNN output. The trust score summarizes the output quality into a single value that can be used for fast decision-making:

\begin{equation}
T_i = \alpha c_i + \beta(1-\sigma_i) + \gamma(1-d_i) + \delta f_i,
\label{eq:trust-score}
\end{equation}

where ($T_i$) is the trust score of output ($i$), and ($\alpha$), ($\beta$), ($\gamma$), and ($\delta$) are non-negative weights. These weights control the importance of confidence, noise, drift, and freshness. They satisfy

\begin{equation}
\alpha + \beta + \gamma + \delta = 1.
\end{equation}

%The trust score increases when confidence and freshness are high, and decreases when noise or drift are high. The weights can be selected according to the application. For a safety-critical task, drift and noise may receive larger weights. For a real-time task, freshness may be more important. In the experimental evaluation, different weight configurations can be tested to analyze their effect on reliability and latency.

The weighted additive form in Eq.~\eqref{eq:trust-score} is intentionally selected for edge-level decision-making. Edge nodes may operate under limited computational resources and cannot always execute expensive probabilistic inference after each physical invocation. The proposed score therefore provides a bounded, monotonic, and interpretable trust estimate that can be computed in constant time when the required telemetry values are available.

\textit{Boundedness.}
If \(c_i, \sigma_i, d_i, f_i \in [0,1]\), and if \(\alpha,\beta,\gamma,\delta \geq 0\) with \(\alpha+\beta+\gamma+\delta=1\), then the trust score \(T_i\) is bounded in the interval \([0,1]\). This follows because the terms \(c_i\), \(1-\sigma_i\), \(1-d_i\), and \(f_i\) are all in \([0,1]\). Since \(T_i\) is a convex combination of these bounded terms, \(T_i\) must also lie in \([0,1]\).

\textit{Monotonicity.}
The score is monotonic with respect to the quality indicators. Increasing confidence \(c_i\) or freshness \(f_i\) increases the trust score, while increasing noise \(\sigma_i\) or drift \(d_i\) decreases the trust score. This behavior is desirable for PNN outputs because the trust score should reward stable, fresh, and confident physical evidence while penalizing uncertainty and substrate degradation.

\subsection{Edge-Level Decision Policy}

The edge layer uses the trust score to decide how the output should be handled. The decision rule is defined as:

\begin{equation}
    D_i =
    \begin{cases}
    \text{Accept}, & T_i \geq \theta_{accept},\\
    \text{Forward to Fog}, & \theta_{reject} \leq T_i < \theta_{accept},\\
    \text{Reject}, & T_i < \theta_{reject},
    \end{cases}
\end{equation}

where ($\theta_{accept}$) and ($\theta_{reject}$) are trust thresholds. If the output has high trust, it is accepted and can be used locally. If the trust score is very low, the output is rejected early to avoid propagating unreliable evidence. If the output is uncertain, it is forwarded to the fog layer. This middle decision region is important because uncertain outputs may still be useful when compared or combined with outputs from other PNNs or edge nodes.

The three-way policy exposes a direct trade-off between local availability and reliability. High-trust outputs can be used immediately at the edge, while clearly unreliable outputs are rejected and intermediate cases are escalated for additional evidence processing. The framework therefore avoids unconditional forwarding, but does not assume that fog escalation is rare. Under challenging uncertainty conditions, a large fraction of outputs may be forwarded intentionally because the objective is to avoid forcing unreliable local decisions. Communication cost itself is not optimized in this work and depends on the selected trust thresholds and operating conditions.
\subsection{Fog-Level Compatibility Checking and Fusion}
\label{sec:fog-fusion}

The fog layer receives uncertain evidence records that have been
forwarded by edge nodes because their edge-level trust scores fall
within the intermediate decision region. Rather than forcing an
independent decision from each uncertain output, the fog layer
compares evidence from multiple PNN sources and performs fusion only
when the evidence is sufficiently compatible and mutually consistent.

Before fusion, the fog layer performs compatibility checking.
In the current framework, two or more evidence records are considered
compatible when they refer to the same task and output modality.
Their timestamps must also fall within a configured temporal window.
When decision-context or observation-window identifiers are available
in the provenance metadata, these identifiers must also be consistent
across the evidence records. Therefore, compatibility is determined
using the task identifier, output modality, timestamp, and available
contextual provenance information. Evidence records that fail these
checks are not fused.

For a set of $N$ compatible numerical outputs
$\{v_1,\ldots,v_N\}$, the fog layer next evaluates the degree of
disagreement among the sources. To reduce sensitivity to an extreme
individual value, the median of the compatible outputs is first used
as a robust reference value:

\begin{equation}
\tilde{v}
=
\operatorname{median}
\left(
v_1,\ldots,v_N
\right).
\label{eq:fog-median}
\end{equation}

The normalized disagreement is then defined as

\begin{equation}
\Delta_{\mathrm{fog}}
=
\frac{
\displaystyle
\max_{i}
\left|
v_i-\tilde{v}
\right|
}{
\max
\left(
|\tilde{v}|,1
\right)
}.
\label{eq:fog-disagreement}
\end{equation}

The denominator prevents numerical instability when the reference
value is close to zero. Fusion is permitted only when

\begin{equation}
\Delta_{\mathrm{fog}}
\leq
\theta_{\mathrm{dis}},
\label{eq:fog-disagreement-condition}
\end{equation}

where $\theta_{\mathrm{dis}}$ denotes the maximum acceptable
disagreement threshold. If the disagreement exceeds this threshold,
the fog layer does not force a fused output. Instead, the current
evidence set is treated as insufficiently consistent, and the system
can request an additional measurement or initiate an appropriate
recovery action, such as recalibration.

When compatibility and disagreement conditions are satisfied, the fog
layer computes a reliability factor for each source. The fog-level
reliability factor is defined as

\begin{equation}
R_i
=
\alpha_f c_i
+
\gamma_f(1-d_i)
+
\delta_f f_i,
\label{eq:fog-reliability}
\end{equation}

where $c_i$ denotes the confidence of source $i$, $d_i$ denotes its
normalized drift, and $f_i$ denotes its freshness. The non-negative
fog-level reliability weights satisfy

\begin{equation}
\alpha_f
+
\gamma_f
+
\delta_f
=
1.
\label{eq:fog-reliability-constraint}
\end{equation}

The fog-level reliability factor $R_i$ is distinct from the
edge-level trust score $T_i$ defined in Eq.~\eqref{eq:trust-score}.
The edge-level trust score combines confidence, uncertainty, drift,
and freshness and is used to select among local acceptance, rejection,
and forwarding. In contrast, $R_i$ is used only during fog-level
fusion. The uncertainty term is deliberately excluded from $R_i$
because uncertainty is incorporated separately through the fusion
weight. This avoids penalizing the same uncertainty estimate once
inside the reliability factor and again through inverse-uncertainty
weighting.

The final fusion weight assigned to source $i$ is defined as

\begin{equation}
w_i
=
\frac{
R_i
}{
\max
\left(
\sigma_i,
\sigma_{\min}
\right)^2
+
\epsilon
},
\label{eq:fusion-weight}
\end{equation}

where $\sigma_i$ denotes the normalized uncertainty associated with
source $i$, $\sigma_{\min}>0$ is a minimum uncertainty floor, and
$\epsilon>0$ is a small constant for numerical stability. The
uncertainty floor prevents a source with a near-zero estimated
uncertainty from receiving an excessively large fusion weight.

The final numerical output is then computed as the normalized weighted
average

\begin{equation}
\hat{y}
=
\frac{
\displaystyle
\sum_{i=1}^{N}
w_i v_i
}{
\displaystyle
\sum_{i=1}^{N}
w_i
}.
\label{eq:fused-result}
\end{equation}

Consequently, a source receives greater influence when it has high
confidence, low drift, good freshness, and low estimated uncertainty.
Confidence, drift, and freshness determine the fog-level reliability
factor $R_i$, while uncertainty affects the source separately through
the inverse-uncertainty component of $w_i$.

The complete fog-level procedure therefore consists of three stages.
First, the system filters evidence according to compatibility.
Second, it evaluates whether the compatible sources agree sufficiently
according to Eq.~\eqref{eq:fog-disagreement}. Third, only evidence that
passes both checks is combined using
Eqs.~\eqref{eq:fog-reliability}--\eqref{eq:fused-result}. Evidence
that is incompatible or exhibits excessive disagreement is not forced
into a fused decision and is instead handled through additional
measurement or recovery mechanisms.

\subsection{Adaptive Reliability Tracking and Recalibration Triggering}
\label{sec:adaptive-trust}

Physical neural substrates may change over time because of drift, aging, environmental variation, biological variability, or calibration loss. Evaluating every invocation independently may therefore fail to detect gradual or persistent degradation. To provide temporal context, the framework maintains a historical trust estimate for each substrate.

Let $\bar{T}_i(t)$ denote the historical trust estimate of substrate $i$ after invocation $t$. It is updated using an exponential moving average:

\begin{equation}
\bar{T}_i(t)
=
\lambda \bar{T}_i(t-1)
+
(1-\lambda)T_i(t),
\label{eq:historical-trust}
\end{equation}

where $\lambda \in [0,1]$ controls the balance between historical and recent observations. A larger $\lambda$ produces a smoother long-term estimate, whereas a smaller value makes the historical trust more responsive to recent changes.

The current implementation combines historical trust degradation with a counter of consecutive low-trust observations. A recalibration request is generated when persistent low current trust or sufficiently low historical trust indicates sustained degradation. This mechanism helps distinguish temporary reliability fluctuations from longer-term changes in substrate behavior. The resulting trigger can be passed to an external substrate-management, calibration, or digital-twin component. The present evaluation validates historical trust tracking and recalibration triggering; it does not perform physical recalibration of a real PNN substrate or automatically update a deployed digital twin.

Algorithm~\ref{alg:trust_management} summarizes the proposed trust-aware output management process.
\begin{algorithm}[!t]
\caption{Trust-Aware Management of PNN Outputs}
\label{alg:trust_management}
\begin{algorithmic}[1]

\Require PNN output $v_i$, confidence $c_i$, uncertainty $\sigma_i$,
drift $d_i$, timestamp $t_i$, modality $m_i$, provenance $\rho_i$

\Ensure Edge-level acceptance or rejection, fog-level fused result,
or uncertain decision

\State Build evidence record
$r_i \gets (v_i,c_i,\sigma_i,d_i,t_i,m_i,\rho_i)$

\State Normalize confidence, uncertainty, drift, and freshness-related values

\If{required reliability telemetry is missing}
    \State Mark the evidence as incomplete
    \State Forward the evidence record $r_i$ to the fog layer
\Else
    \State Compute freshness $f_i$ using Eq.~\eqref{eq:freshness}
    \State Compute edge-level trust score $T_i$ using Eq.~\eqref{eq:trust-score}

    \If{$T_i \geq \theta_{\mathrm{accept}}$}
        \State Accept the PNN output at the edge
        \State Update historical trust for substrate $i$
        \State \Return $v_i$

    \ElsIf{$T_i < \theta_{\mathrm{reject}}$}
        \State Reject the PNN output as unreliable
        \State Update historical trust for substrate $i$
        \State \Return rejection decision

    \Else
        \State Forward the evidence record $r_i$ to the fog layer
    \EndIf
\EndIf

\State Identify evidence records compatible in task identifier,
modality, timestamp window, and available contextual provenance

\If{no compatible evidence set exists}
    \State Request additional evidence or measurement
    \State Update historical trust for the involved substrate(s)
    \State \Return uncertain decision
\EndIf

\State Compute disagreement $\Delta_{\mathrm{fog}}$
using Eq.~\eqref{eq:fog-disagreement}

\If{$\Delta_{\mathrm{fog}} > \theta_{\mathrm{dis}}$}
    \State Request additional measurement or initiate recovery action
    \State Update historical trust for the involved substrate(s)
    \State \Return uncertain decision
\EndIf

\For{each compatible evidence record $r_j$}
    \State Compute fog-level reliability factor $R_j$
    using Eq.~\eqref{eq:fog-reliability}

    \State Compute fusion weight $w_j$
    using Eq.~\eqref{eq:fusion-weight}
\EndFor

\State Compute fog-level fused result $\hat{y}$
using Eq.~\eqref{eq:fused-result}

\State Update historical trust for the involved substrate(s)

\State \Return $\hat{y}$

\end{algorithmic}
\end{algorithm}

\subsection{Computational Complexity}
\label{sec:complexity}

The proposed framework keeps the latency-sensitive edge processing lightweight. Evidence construction, missing-telemetry checking, freshness computation, trust-score evaluation, and the three-way decision rule each require $O(1)$ time per output when the required quality indicators are already available. If uncertainty is estimated from $K$ repeated measurements, the corresponding edge-level cost becomes $O(K)$ per source. The adaptive trust update also requires only $O(1)$ time per invocation.

At the fog layer, compatibility checking and numerical fusion over $N$ evidence records require $O(N)$ time, while the median-based disagreement analysis requires $O(N \log N)$ in the current implementation. Therefore, the overall complexity is dominated by disagreement checking and is $O(N \log N)$ when quality indicators are precomputed. When uncertainty estimation from $K$ repeated measurements is included for all $N$ sources, the end-to-end worst-case complexity becomes $O(NK + N \log N)$.

\section{Evaluation \& Analysis}\label{sec:eval}
\subsection{Experimental Setup and Metrics}
\label{sec:experimental-setup}

The evaluation examines whether explicit post-invocation reliability management improves the handling of PNN outputs after successful substrate invocation. The proposed framework is implemented as an extension of the \cpntwo~\cite{fischer2026phys} prototype. The underlying \cpntwo workflow remains responsible for substrate discovery, matching, invocation, and telemetry access, whereas the proposed layer operates on the returned output and its associated reliability information.

The experiments use \cpntwo-compatible numerical PNN output models under controlled and randomized uncertainty conditions. Hidden physical noise and drift are used to generate corrupted outputs, while the trust layer receives imperfect estimates of these quantities rather than their exact ground-truth values. Confidence is generated from an independent latent backend-quality component together with smaller contributions from estimated noise and drift. Delay influences the physical error through simulated state evolution and independently affects the freshness component of the trust score.

The default edge configuration uses equal trust weights and decision thresholds of $\theta_{\mathrm{accept}} = 0.75$ and $\theta_{\mathrm{reject}} = 0.35$. Equal weights are used as a neutral reference configuration rather than as an assumed optimal setting. Separate sensitivity experiments examine alternative weight priorities and different levels of decision-policy strictness.

The edge-level evaluation compares four methods: direct use of the raw \cpntwo output, confidence-only selection, freshness-only selection, and the proposed full trust score. The controlled evaluation varies physical noise, drift, output delay, and the freshness time constant $\tau$. A separate randomized stress experiment samples a broader range of uncertainty conditions. The main controlled, randomized, and fog-fusion experiments are repeated across 20 independent random seeds to evaluate statistical stability.

We distinguish several reliability quantities that capture different aspects of decision quality. \emph{Coverage} is defined as $P(\mathrm{Accept})$. The \emph{conditional false acceptance rate} (FAR) is defined as $P(\mathrm{Accept}\mid\mathrm{Bad})$, while the \emph{unsafe acceptance frequency} is defined as $P(\mathrm{Accept}\cap\mathrm{Bad})$. The \emph{accepted-output contamination rate} is defined as $P(\mathrm{Bad}\mid\mathrm{Accept})$. These metrics have different denominators and therefore should not be interpreted interchangeably. The conditional false rejection rate is similarly defined as $P(\mathrm{Reject}\mid\mathrm{Good})$. Accepted-output mean absolute error (MAE) measures the numerical error only among outputs that are accepted locally.

Because a lower acceptance rate can itself reduce accepted-output error, a separate risk--coverage experiment varies the acceptance threshold while keeping the rejection threshold fixed. This experiment evaluates whether the proposed method continues to achieve lower selective risk than confidence-only filtering at comparable coverage levels.

At the fog layer, seven numerical fusion methods are evaluated: simple averaging, confidence weighting, inverse-uncertainty weighting, trust-only weighting, median fusion, the original $T_i/\sigma_i^2$ formulation, and the proposed orthogonal $R_i/\sigma_i^2$ formulation. Compatibility and normalized disagreement are assessed separately from numerical fusion. Deterministic functional tests additionally verify the rejection of evidence with mismatched task identifiers, modalities, decision contexts, observation windows, or timestamps.

Additional experiments evaluate trust-component ablation, missing reliability telemetry, adaptive historical trust, and computational scalability. Unless otherwise stated, confidence intervals in the multi-seed evaluation represent 95\% confidence intervals computed over seed-level aggregate results.

\subsection{Results and Discussion}
\label{sec:results-discussion}

Figure~\ref{fig:edge-delay} examines accepted-output quality as delay increases. \cpntwo and confidence-only continue to accept outputs even at long delays because neither policy explicitly prevents stale evidence from being used. Freshness-only stops accepting once the freshness value becomes too low, while the full trust method also stops local acceptance when the combined evidence is insufficiently reliable. Consequently, their curves end because accepted-output MAE is undefined when no outputs are accepted locally; the missing points therefore represent selective forwarding or rejection, not zero error or missing experimental data. The non-monotonic MAE of the remaining methods is expected because MAE is computed only over the subset accepted at each delay and the physical error also depends on the simulated state evolution.

\inlinefigpage
{1}
{Accepted-output MAE as delay increases. Shorter curves indicate operating regions in which a policy accepts no outputs locally, making accepted-output MAE undefined.}
{fig:edge-delay}

Figure~\ref{fig:risk-coverage} examines whether the lower accepted-output error of the full trust method results only from accepting fewer outputs. Over the overlapping coverage range, the full-trust curve remains below the confidence-only curve, showing that the multi-factor trust score ranks reliable outputs more effectively than confidence alone. \cpntwo appears as a single reference point rather than a curve because it is an accept-all baseline: it has no post-invocation acceptance threshold to vary and therefore operates only at full coverage. The result consequently reflects a reliability--coverage trade-off rather than a comparison based only on one fixed threshold.

\inlinefigpage
{2}
{Reliability--coverage trade-off for confidence-only and full-trust selection. \cpntwo is shown as a single accept-all reference point because it has no post-invocation reliability threshold to sweep.}
{fig:risk-coverage}

Figure~\ref{fig:coverage-drift} shows the availability cost of responding to substrate degradation. \cpntwo maintains full local acceptance because it performs no reliability filtering. Confidence-only gradually reduces acceptance, whereas full trust reduces it more strongly because increasing drift directly lowers the trust score. The falling full-trust curve therefore represents deliberate escalation of degraded evidence rather than loss of functionality.

\inlinefigpage
{3}
{Local acceptance under increasing substrate drift. Full trust progressively shifts uncertain outputs away from local acceptance as degradation becomes stronger.}
{fig:coverage-drift}

Figure~\ref{fig:coverage-noise} shows a similar reliability--availability response to physical noise. \cpntwo continues accepting all outputs, while confidence-only becomes more selective and full trust reduces local acceptance more rapidly. This behavior is consistent with the proposed policy: strong uncertainty is handled through forwarding or rejection instead of preserving edge coverage at any cost.

\inlinefigpage
{4}
{Local acceptance under increasing physical noise. The full trust policy becomes more selective as output uncertainty increases.}
{fig:coverage-noise}

Figure~\ref{fig:unsafe-drift} shows the safety consequence of these decisions. Unsafe acceptance increases strongly for raw \cpntwo as drift grows, whereas confidence-only remains substantially lower. Full trust maintains the lowest unsafe acceptance and approaches zero under severe drift. This reduction must be interpreted together with Figure~\ref{fig:coverage-drift}: part of the safety gain results from deliberately withholding unreliable outputs from local acceptance.

\inlinefigpage
{5}
{Safety and local availability under increasing physical uncertainty at $\tau=5$~s.}
{fig:unsafe-drift}

Figure~\ref{fig:unsafe-noise} shows the same pattern under physical noise. Raw \cpntwo increasingly propagates unreliable outputs, while confidence-only reduces the risk and full trust suppresses it further. At high noise levels, the near-zero unsafe acceptance of full trust coincides with very low local coverage, demonstrating the intended safety--availability trade-off rather than claiming that highly corrupted outputs become accurate.

\inlinefigpage
{6}
{Safety and local availability under increasing physical uncertainty at $\tau=5$~s.}
{fig:unsafe-noise}

Figure~\ref{fig:ablation-noise} shows how the individual trust components affect accepted-output quality under increasing noise. Removing uncertainty has the clearest effect at severe noise because the policy loses the component that directly represents short-term physical instability. Removing drift also degrades accepted-output quality because the noise bins still contain scenarios with different substrate conditions. Some curves terminate at higher noise levels because the corresponding variants accept no outputs in those bins; their missing MAE values therefore indicate selective behavior rather than zero error.
\inlinefigpage
{7}
{Accepted-output MAE under increasing noise for the complete trust model and each ablated variant.}
{fig:ablation-noise}

Figure~\ref{fig:ablation-drift} isolates the role of drift information. The complete method increasingly suppresses unsafe local acceptance as degradation becomes stronger. When drift is removed, unsafe acceptance no longer falls in the same way and remains present across degraded conditions because the policy cannot directly recognize movement away from calibrated behavior. Other components can still reduce acceptance indirectly, but they do not replace an explicit drift signal.

\inlinefigpage
{8}
{False acceptance under increasing substrate drift for the full trust model and its component ablations.}
{fig:ablation-drift}

Figure~\ref{fig:ablation-delay} demonstrates the distinct role of freshness. Variants that retain freshness rapidly stop unsafe local acceptance as evidence ages. Removing freshness leaves a persistent unsafe-acceptance level even at long delays, because confidence, uncertainty, and drift contain no direct information about the age of the evidence. Freshness therefore contributes temporal validity that cannot be fully replaced by the other trust components.

\inlinefigpage
{9}
{False acceptance under increasing output delay, highlighting the role of freshness in preventing stale-output acceptance.}
{fig:ablation-delay}

Figure~\ref{fig:fusion-sources} compares numerical fusion as the number of PNN sources increases. All evaluated methods show lower mean error with additional sources, indicating the benefit of combining more independent evidence. Simple averaging, confidence weighting, and median fusion remain weaker because they do not jointly account for physical uncertainty and reliability. The uncertainty-aware methods form the strongest group. The proposed $R_i/\sigma_i^2$ formulation remains very close to the uncertainty-only and original trust--uncertainty approaches and achieves the lowest aggregate mean error across the 20-seed evaluation, although the numerical advantage is small. The main contribution of the fog layer should therefore be interpreted as the combination of compatibility checking, disagreement gating, and reliability-aware fusion rather than a large improvement from weighting alone.

\inlinefigpage
{10}
{Mean fusion error as the number of PNN sources increases for the seven evaluated fusion strategies.}
{fig:fusion-sources}

Figure~\ref{fig:fusion-availability} shows that disagreement gating becomes more selective as the evidence set grows. With more sources, the probability that at least one value exceeds the maximum allowed deviation from the median increases, so fewer evidence sets pass the fixed disagreement threshold. This reduction in fusion availability is therefore an expected consequence of the gating rule rather than evidence that numerical fusion becomes less accurate with additional sources.

\inlinefigpage
{11}
{Fusion-allowed rate as the number of PNN sources increases under the fixed disagreement threshold.}
{fig:fusion-availability}

Figure~\ref{fig:adaptive-trust} illustrates how the adaptive mechanism responds to persistent changes in substrate reliability. Current trust reacts quickly to individual observations, whereas the exponential moving average changes more slowly and filters short-term variation. During sustained degradation, repeated low-trust observations eventually satisfy the persistence condition and generate a recalibration request. During the simulated recovery phase, current trust improves first while historical trust follows more gradually. The experiment demonstrates the intended temporal behavior of the trigger, but does not represent physical recalibration or automatic digital-twin synchronization.

\inlinefigpage
{12}
{Adaptive current and historical trust during healthy operation, sustained degradation, and simulated recovery.}
{fig:adaptive-trust}

\subsection{Scalability and Limitations}
\label{sec:scalability-limitations}

Figure~\ref{fig:edge-scalability} shows that the total edge-level processing time increases approximately proportionally to the number of evidence records. The corresponding per record measurements remain close to $2\,\mu\mathrm{s}$ across the tested workload sizes, which is consistent with the $O(1)$ trust-score and decision cost per record when quality indicators are already available.

\inlinefigpage
{13}
{Edge-level trust scalability. Total runtime grows approximately proportionally with the number of processed evidence records.}
{fig:edge-scalability}

Figure~\ref{fig:fog-scalability} shows that both numerical fusion and compatibility/disagreement assessment remain lightweight for the evaluated source counts. Their measured runtime increases smoothly with the number of sources. Numerical fusion is $O(N)$, while the current median-based disagreement procedure gives the complete assessment an $O(N\log N)$ worst-case complexity. The fact that compatibility/disagreement assessment is faster than numerical fusion in the tested range does not contradict this asymptotic result; for the evaluated source counts, implementation constants dominate the theoretical growth-rate difference.

\inlinefigpage
{14}
{Fog-level processing scalability for numerical fusion and compatibility/disagreement assessment as the number of PNN sources increases.}
{fig:fog-scalability}

The additional validation experiments support the same interpretation. Under the deliberately harsh randomized stress distribution, the full trust policy becomes substantially more conservative than confidence-only filtering and maintains markedly lower unsafe acceptance and accepted-output error. The sensitivity analysis confirms that stricter decision thresholds reduce unsafe acceptance at the cost of lower local coverage, while different weight priorities shift the balance among physical failure modes. Missing-telemetry tests show that outputs lacking required noise, drift, or freshness information are conservatively forwarded rather than locally accepted. Finally, all deterministic fog-compatibility tests correctly reject mismatched task, modality, decision-context, observation-window, and timestamp conditions.

\section{Conclusion \& Future Work}
\label{sec:conclusion}

This paper introduced a trust-aware framework for managing PNN outputs after invocation in cloud-continuum systems. Instead of treating every returned result as reliable, the framework evaluates output quality at the edge, forwards uncertain cases to the fog, and applies compatibility checking, disagreement screening, and reliability-aware fusion when multiple PNN outputs are available.

The results show a clear improvement in output reliability. Across 20 random seeds, unsafe acceptance decreased from about 61.5\% for raw \cpntwo\ output handling to about 5.0\%, while accepted-output MAE decreased from approximately 0.504 to 0.242. The risk--coverage analysis also shows that this improvement is not only caused by lower local acceptance, since the full trust method performs better than confidence-only filtering at similar coverage levels. At the fog layer, the proposed fusion method provides a small but consistent improvement over strong uncertainty-aware baselines, while the main benefit comes from combining compatibility, disagreement, and reliability checks before fusion.

The current prototype mainly relies on controlled and synthetic uncertainty scenarios. Future work will focus on trace-driven and real PNN experiments, where confidence, uncertainty, and drift can be estimated directly from substrate behavior. Further work will also study adaptive thresholds, disagreement sensitivity, and tighter integration with recalibration and digital-twin management.

\section{Data Availability}
\label{Data Availability}
\url{https://github.com/MaliheHa93/Trust_PNNs}

% Bibliography inlined for arXiv portability
\bibliographystyle{model1-num-names}
\bibliography{cas-refs}

% Author biographies: preserve the original IEEE Access layout.
% Do not force a page/column break here: IEEEbiography lets the references
% continue in the left column while the biographies begin in the right column.
\begin{IEEEbiography}[{\includegraphics[width=1in,height=1.25in,clip,keepaspectratio]{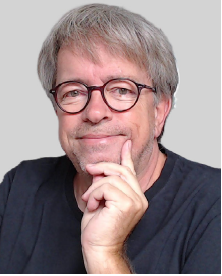}}]{Stefan Fischer} is a full professor of computer science at the University of L\"ubeck, Germany, and the Director of the Institute of Telematics. He received the Diploma degree in information systems and the doctoral degree in computer science from the University of Mannheim, Germany, in 1992 and 1996, respectively. After a postdoctoral year at the University of Montreal, Canada, he held positions at the International University in Germany and the Technical University of Braunschweig before joining the University of L\"ubeck in 2004. His research interests include distributed systems, sensor and ad hoc networks, the Internet of Things, smart cities, molecular communications, and physical neural computing.
\end{IEEEbiography}

\vspace{0.5\baselineskip}

\begin{IEEEbiography}[{\includegraphics[width=1in,height=1.25in,clip,keepaspectratio]{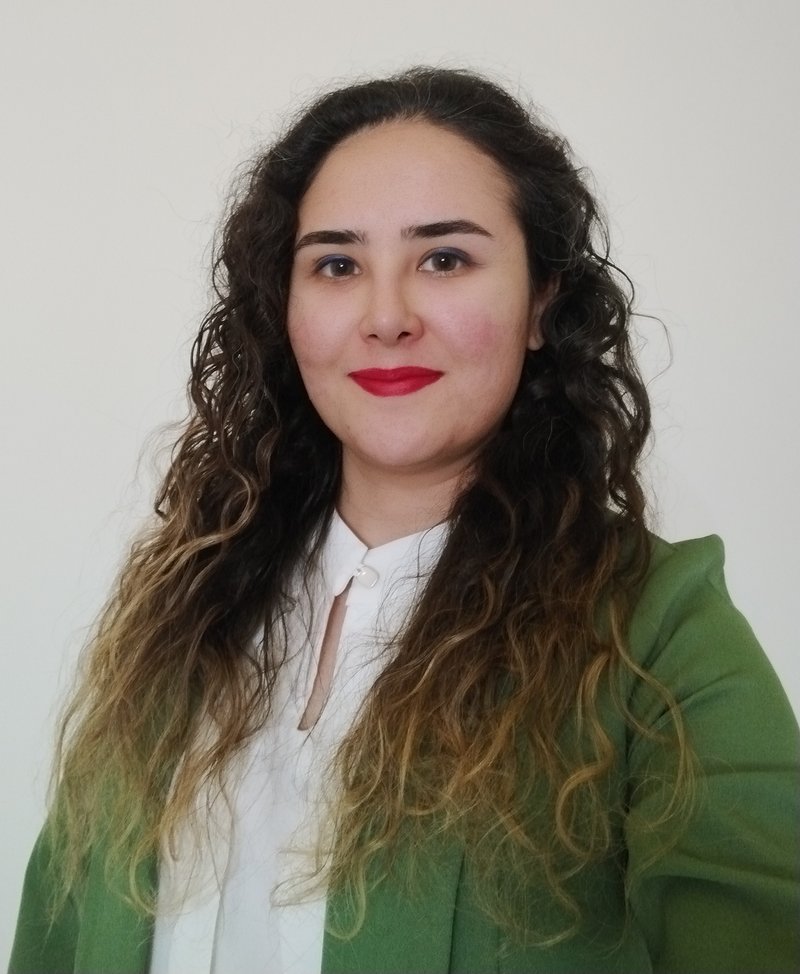}}]{Maliheh Hariri}
 is a Ph.D. student in computer science at the University of L\"ubeck, Germany, working on distributed intelligent systems, edge and fog computing, and emerging computing paradigms, especially in scheduling and resource management. Currently, her research focuses on the integration of Physical Neural Networks into cloud-continuum environments. She has also been working on scientific workflow scheduling, Service Function Chain placement, and resource optimization in fog and cloud computing systems. Her broader research interests include physical AI, edge intelligence, distributed AI infrastructures, and reliable decision-making under uncertainty.
\end{IEEEbiography}

\EOD
\end{document}